\documentclass[twocolumn,showpacs,showkeys,preprintnumbers,amsmath,amssymb]{revtex4}
\usepackage[T2A]{fontenc}
\usepackage[cp1251]{inputenc}
\usepackage[russian,english]{babel}
\usepackage{array,amsmath,amstext,amssymb}
\usepackage{graphicx,color}
\begin{document}
\title{REFLEXIVE SPATIAL BEHAVIOUR DOES NOT GUARANTEE EVOLUTION ADVANTAGE IN PREY--PREDATOR COMMUNITIES}
\author{Michael G.\,Sadovsky}
\affiliation{Institute of computational modelling of SD of RAS;\\ 660036 Russia, Krasnoyarsk.}
\email{msad@icm.krasn.ru}

\author{Maria Yu.\,Senashova}
\affiliation{Institute of computational modelling of SD of RAS;\\ 660036 Russia, Krasnoyarsk.}
\email{msen@icm.krasn.ru} 

\newcounter{N}
\begin{abstract}
We consider the model of spatially distributed population consisting of two species with ``\textsl{predator\,--\,prey}'' interaction;  each of the species occupies two stations. Transfer of individuals between the stations (migration) is not random and yields the maximization of a net reproduction of each species. Besides, each species implements reflexive behavior strategy to determine the optimal migration flow.
\end{abstract}

\pacs{87.23.Gc}

\keywords{interaction, targeted migration, dynamics, spatial distribution, non-diffusive, information access}

\maketitle

\section{Introduction}
Modelling of the dynamics of biological communities is of great importance. The pioneering fundamental works \cite{volt,volt1} started the development of mathematical modelling of the dynamics of various biological populations (see also \cite{lotka,turch,petrovsky,Wys_Hust,morozov,japy}). Here populations are supposed to be a kind of chemical reactor where various ``chemical'' reactions run (namely, reproduction, elimination and other types of interactions of organisms).

Studying spatially distributed communities, one faces the stunning prevalence of the papers based on ``reaction~-- diffusion'' ideology. Nonetheless, a coincidence of a solution of such system and real dynamics usually brings a kind of misconception: a diversity and abundance of the (structurally stable) solutions of such ``reaction~-- diffusion'' systems exceeds any really available trajectory families. Thus, one always is able to figure out PDE (or ODE) with the behaviour pretty close to an observed one. The key issue is that all the organisms, including microorganisms, differ in their microscale behaviour\footnote{Microscale behaviour is the spatial behaviour executed at the distances comparable to the size of an organism.} from the chemical compounds and relative chemical reactions.

Vito Volterra, the founder of mathematical population biology, was very keen in this point, when implemented a chemically based equation system for a description of a community behaviour. Yet, the ``reaction~-- diffusion'' systems make the basis for modelling of spatially distributed populations and communities \cite{Wys_Hust,petrovsky,Strohm_Tyson,invasion,waves,morozov,japy,PanPing,ZijianL,JianjunJ,ZijuanW,Dhar,JianjunJ1,Kiss}. This approach puts on a strong constraint on the properties of organisms under consideration:
\begin{list}{--}{\topsep=0mm \labelsep=2mm \itemsep=0mm \parsep=0mm \itemindent=3.02mm}
\item organisms must transfer randomly and aimlessly;
\item organisms must have no memory, and finally,
\item the greatest majority of the transfers in space must be of a small (or even infinitesimally small) scale.
\end{list}
Obviously, none of these constraints hold true (see, e.\,g., \cite{sci2000,insect,annals}) even for microorganisms \cite{gorban,zelkniga,levitt,ameba}.

Another problem of the ``reaction~-- diffusion'' approach is that long distance transfers have extremely low probability~$\sim \exp\{-\l^2\}$ (here~$l$ is a typical distance of a transfer). A diffusion, in such capacity, means that changing a habitation site for another one (located infinitesimally close to the original one), a being has to arrange \textsl{de novo} a nest (or any other type of habitation). Yet, any significant change in environmental conditions never take place at the small distances. This fact follows in a loss of any advantages from a transfer; just contrary, a small scale transfer would result in the increased senseless wastes of all the vital resources (time, efforts, increased risk of an outer attack, etc.), while a change of environmental conditions would be infinitesimally small. Thus, a migration makes sense only when it is long enough. This fact makes a change of continuous spatial pattern for a discrete one quite natural and clear.

Essentially new class of model must be implemented to overcome those discrepancies. The core principle of these models is that a transfer of beings in space must meet the evolution optimality conditions and result in a growth of a population viability\footnote{We shall not get into a long standing discussion of what is an issue of this viability improvement. }. To get that, beings must transfer in space smartly; it means that a transfer must bring a maximization of net reproduction (an average off-spring number \textsl{per capita} determined for a sufficiently long line of generations), see details in~\cite{p5,p6,p7,gorban}.

Here we consider the model of a two-species spatially distributed community, where the species interact following the ``\textsl{prey~--~predator}'' pattern. Both species are split on two subpopulations each occupying a station. Migration is a transfer from one station to another; no other movements in space (ordinary in a real situation) are assumed to affect a community dynamics. Both species (in various combinations) are supposed to implement a reflexive behaviour.

\subsection{Conflicts and dynamics of communities}
Modeling of a behaviour where few agents are to maximize the same payoff function is rather sounding in a variety of science fields ranging from sociology, ethology, psychology or environmental sciences \cite{m1,m2,m3}) to quite theoretical issues in the theory of optimal control and game theory \cite{m4,m5}). A comprehensive study of the dynamics of such systems requires a straight and unambiguous problem formulation. The formulation, in turn, is mainly determined by a specific system under consideration where a conflict interaction (or competition) takes place.

That was Vladimir Lefevre \cite{p2,p3,p4} who pioneered the studies of a conflict behaviour. A wide spread of mathematical modelling ideology over the studies of biological systems surprisingly brought very few efforts in implementation of a conflict behaviour into the biological systems dynamics studies \cite{m1,m2,m3,m4,m5}. A study of the dynamics of a spatially distributed biological community is a good point to apply the conflict analysis techniques. In general, an impact of a spatial patterns on the dynamics of biological communities makes a complicated and nontrivial problem.

In this paper, we present an approach to describe a conflict behaviour of a spatially distributed biological community including those dynamical peculiarities which result from the targeted and smart behaviour of agents, in a conflict. We shall concentrate on a study of classical two-species \textsl{prey~-- predator} community. It means that the organisms of the first species (\textsl{preys}) exist due to an external resource, but the organisms of another species (\textsl{predator}) have a solely source of a food from the predation of those preys; this system is a classical object of a study in mathematical ecology \cite{p5}.

\section{Models of spatial behaviour of two species population}
We consider community consisting of two species with \textsl{prey~-- predator} interaction. We assume that each of the
populations from the community is split into two subpopulations occupying two separated habitats (stations). A migration is the transfer from station to station, only. Any movement of individuals within a station are neglected. No effects on the population dynamics from a spatial pattern are presumed, for each subpopulation, when no migration takes place. The dynamics of a community will be studied in discrete time; a study in a continuous time is possible, as well, while it brings nothing essential but the tremendous technical problems; see also \cite{lattice}.

\subsection{The basic model of migration}\label{bazmod}
Further, $N_t$ ($M_t$, respectively) is the abundance of the prey subpopulation in station \fbox{I} (in station \fbox{II}\,, respectively); similarly,
$X_t$ и $Y_t$ is the abundance of the predator subpopulation in those stations. We suppose the dynamics of a subpopulation in each station (in the migration-free case) to follow the discrete version of Lotka-Volterra classical equation\footnote{Some newly results extending this approach could be found in \cite{Wys_Hust}; technically advanced results in this type of classical model could be found in \cite{petrovsky}.} \cite{p5,p7,l}:
\begin{subequations}\label{eq1}
\renewcommand{\theequation}{\theparentequation\alph{equation}}
\begin{equation}\label{eq1:1}
\begin{array}{rcl}
N_{t+1}&=&N_t \cdot \left(a - bN_t - fX_t\right)\\ X_{t+1}&=&X_t \cdot \left(\varepsilon fN_t -
hX_t\right)
\end{array}
\end{equation}
\begin{equation}\label{eq1:2}
\begin{array}{rcl}
M_{t+1}&=&M_t \cdot \left(c - dM_t - gY_t\right)\\
Y_{t+1}&=&Y_t\cdot \left(\varepsilon gM_t - kY_t\right)\,.
\end{array}
\end{equation}
\end{subequations}
Here  $a$ and $c$ are prey fertility,  $b$ and $d$ describe the density-dependent intraspecific competition among the preys,  $f$ and $g$ describe the interactions between predators and prey, and  $h$ and $k$ describe density-dependent intraspecific competition among predators in the first and second station, respectively. These parameters determine (within the framework of the model, of course) the ecological capacity of a station: the figure~$\sim b^{-1}$ ($h^{-1}$, respectively) makes a specific (average) subpopulation abundance existing within a station. The parameter $\varepsilon$ describe the transformation efficiency of prey biomass into predator biomass (assimilation coefficient). The terms $\varepsilon f N$ and $\varepsilon g M$ are similar to parameters $a$ and $c$ (fertility) for preys (they depending on the prey abundance).

Next, we shall suppose that a transfer between stations may cause the losses of individuals of both species due to outer reasons or due to
interspecific interactions. These losses will be characterized by \textbf{migration cost} figures $p$, $0<p\leqslant 1$ for preys and $q$, $0<q\leqslant 1$ for predators, respectively. A possible interpretation of the migration cost is the probability of successful migration from one station to another; success here means that no damage for further reproduction occurs. Functions in the parentheses in \eqref{eq1} are \textbf{net reproduction} figures (NR) of the corresponding subpopulations.

The migration of each species runs in a manner to maximize the average NR over two stations; individuals of each species migrate independently.
A migration at time step $t$ from station~\textbf{I} to station~\textbf{II} starts, if living conditions ``there'' are better, than ``here'', with respect to the transfer cost:
\begin{equation}\label{eq2}
\begin{array}{rcl}
\left(a - bN_t - fX_t\right)&<&p \cdot \left(c - dM_t - gY_t\right)\,,\\ \left(\varepsilon fN_t -
hX_t\right)
&<&q \cdot \left(\varepsilon gM_t - kY_t\right)\,.\\
\end{array}
\end{equation}
for preys and predators beings, respectively.  The backward migration conditions are defined similarly:
\begin{equation}\label{eq2_2}
\begin{array}{rcl}
p \cdot \left (a - bN_t - fX_t\right)&>&\left(c - dM_t - gY_t\right)\,,\\ q\cdot \left(\varepsilon
fN_t -
hX_t\right)&>&\left(\varepsilon gM_t - kY_t\right)\,.\\
\end{array}
\end{equation}
Note, that the migration act is independent for each being, while the model considers it as a population event.

If neither of the inequalities (\ref{eq2}, \ref{eq2_2}) hold true, then no migration takes place, at the given time moment~$t$. Prey migration flux $\Delta$ (predator migration flux $\Theta$, respectively) must equalize the inequalities~\eqref{eq2} or~\eqref{eq2_2}:
\begin{subequations}\label{eq3}
\renewcommand{\theequation}{\theparentequation\alph{equation}}
\begin{equation}\label{eq3:1}
\begin{array}{rcl}
\left(a - b(N_t - \Delta) - fX_t\right)&=&p \cdot \left(c - d(M_t+p\Delta) - gY_t\right)\,,\\
\left(\varepsilon fN_t - h(X_t-\Theta)\right)&=&q \cdot \left(\varepsilon gM_t - k(Y_t+q\Theta)\right)\\
\end{array}
\end{equation}
\textrm{for the case (\ref{eq2}), or}
\begin{equation}\label{eq3:2}
\begin{array}{rcl}
p\cdot \left (a - b(N_t + p \Delta) - fX_t\right)&=&\left(c - d(M_t - \Delta) - gY_t\right)\,,\\
q \cdot \left(\varepsilon fN_t - h(X_t+ q \Theta)\right)&=&\left(\varepsilon gM_t - k(Y_t - \Theta)\right)\\
\end{array}
\end{equation}
\end{subequations}
for the case (\ref{eq2_2}). Indeed, the migration flow (defined according to~(\ref{eq4})) can not exceed the total abundance of the subpopulation in appropriate emigration station. If it happens, the migration flow is just equal to the abundance at the respective station. Migration fluxes $\Delta$ and $\Theta$ are equal to
\begin{subequations}\label{eq4}
\renewcommand{\theequation}{\theparentequation\alph{equation}}
\begin{equation}\label{eq4:1}
\begin{cases}
\Delta = \min\left\{N_t, \dfrac{pc-a+bN_t-pdM_t+fX_t-pgY_t}{b+p^2d}\right\}\\[12pt] \Theta = \min\left\{X_t, \dfrac{hX_t+ \varepsilon qgM_t - \varepsilon fN_t -qkY_t}{h+q^2k}\right\}
\end{cases}
\end{equation}
\textrm{for migration from station \fbox{I} to station \fbox{II}, and }
\begin{equation}\label{eq4:2}
\begin{cases}
\Delta = \min\left\{M_t, \dfrac{pa-c+dM_t-pbN_t+gY_t-pfX_t}{d+p^2b}\right\}\\[12pt] \Theta = \min\left\{Y_t, \dfrac{kY_t+ \varepsilon qfN_t - \varepsilon gM_t -qhX_t}{k+q^2h}\right\}\,.
\end{cases}
\end{equation}
\end{subequations}
for the backward migration. Once again, the Eqs.~(\ref{eq4:1}, \ref{eq4:2}) show the case of the coherent migration of both species (both preys and predators migrate the same direction); surely, they might migrate in opposite directions, either.

Finally, let's outline how the basic model (\ref{eq1}~-- \ref{eq4}) works. For each time moment~$t$, a direction and the migration fluxes ($\Delta$ and $\Theta$, respectively) are determined. Then, the species redistribute themselves according to the Eqs.\,(\ref{eq4}) upgrading the abundances in the stations. Then, the next generation numbers $\{N_{t+1}, X_{t+1};\ M_{t+1}, Y_{t+1}\}$ are calculated, according to~(\ref{eq1}), with the upgraded abundances of the current generation $\{\widetilde{N}_{t}, \widetilde{X}_{t};\ \widetilde{M}_{t}, \widetilde{Y}_{t}\}$ defined by (\ref{eq3}). If no
migration takes place at the time moment~$t$, then the stage with individuals redistribution is omitted.

Note that the model (\ref {eq1}~-- \ref {eq4}) (and its modifications shown below) does not guarantee maximization of NR for each species at each step of time (one act of migration), on contrary to the model of a single population dynamics \cite{p5,p6,p7}.

\subsection{A modified model of reflexive behaviour}\label{prin}
Here we  describe the model of the \textbf{reflexive behaviour}. Reflection here means a behaviour accompanied with an ability to foresee and/\!or predict the behaviour of an opponent, in a competitive behavioural act. Basic approaches to describing and modeling such behavior were developed by V.~Lefebvre \cite{p2,p3,p4}. An implementation of reflexive behavioural strategy among animals is well known. Not discussing here psychological or ethological aspects of those strategies, let concentrate on a simple model revealing the dynamic effects of these latter. Obviously, the reflexive behavior of animals is not intelligent, at least, always. In general, rationality and reflexivity are not the synonyms and can be implemented independently.

Three reflexive behavior patterns are possible:
\begin{list}{\textsl{\roman{N}})}{\usecounter{N}\topsep=0mm \labelsep=2mm \itemsep=0mm \parsep=0mm \itemindent=3.02mm}
\item prey reflects the predator behavior;
\item predator reflects the prey behavior; and finally,
\item both species reflect each other behaviors.
\end{list}

Consider the first version: \textbf{prey reflects the predator behavior} with the relevant modification of the basic model. Change the basic model (\ref{eq1}~-- \ref{eq4}) to investigate this effect. Reflexive behavior means that  prey individuals are able ``to predict'' the behaviour of predators; besides, they actively use this knowledge when making the choice of their behavioural pattern. It is expressed through the change of fluxes figures, in our modification;~$\Theta$ is still determined by~(\ref{eq4}), while~$\Delta $ is now determined by:
\begin{equation}\label{eq5}
\begin{array}{rcl}
\Delta&=&\displaystyle\frac{pa-c+dM-pbN+g\widetilde{Y}-pf\widetilde{X}}{d+p^2b} \qquad \textrm{or}\\[5mm]
\Delta&=&\displaystyle\frac{pc-a+bN-pdM+f\widetilde{X}-pg\widetilde{Y}}{b+p^2d}\,,
\end{array}
\end{equation}
where $\widetilde{X} = X + q \Theta$, $\widetilde{Y} = Y - \Theta$ or $\widetilde{X} = X -\Theta$, $\widetilde{Y} = Y + q\Theta$, depending on the direction of migration of predators.

If \textbf{predators reflect the preys behaviour} then~$\Delta$ figures remain the same as at the basic model, but~$\Theta $ figures are determined by:
\begin{equation}\label{eq6}
\begin{array}{rcl}
\Theta&=&\displaystyle\frac{hX+ \varepsilon qg\widetilde{M} - \varepsilon f\widetilde{N} -qkY}{h+q^2k}
\qquad \textrm{or}\\[5mm]
\Theta&=&\displaystyle\frac{kY+ \varepsilon qf\widetilde{N} - \varepsilon g\widetilde{M} -qhX}{k+q^2h}\,,
\end{array}
\end{equation}
where $\widetilde{N} = N + p \Delta$, $\widetilde{M} = M - \Delta$ or $\widetilde{N} = N -\Delta$, $\widetilde{M} = M + p\Delta$, depending on the direction of migration of preys.

Finally, the following model modification presents the case where \textbf{both species reflect each other}. To describe it, arrange the model into a series of consequent steps.
\begin{list}{\arabic{N}.\,}{\usecounter{N}\topsep=0mm \labelsep=2mm \itemsep=0mm \parsep=0mm \itemindent=3.02mm}
\item The directions of migration and the figures of $\Delta$ and $\Theta$ are determined according to the basic model (\ref{eq1}~-- \ref{eq4}).
\item The upgraded abundances $\{\widetilde{N}, \widetilde{M};\ \widetilde{X}, \widetilde{Y}\}$ are determined. Again, here each species determine the upgraded abundances of the ``opponent'' as if that latter does not realize a reflexive strategy.\label{refff}
\item Both\label{ref1111} the directions and fluxes figures $\widehat{\Delta}$ and $\widehat{\Theta}$ (if any) are re-determined according to upgraded values $\{\widetilde{N}, \widetilde{M};\ \widetilde{X}, \widetilde{Y}\}$ calculated at the step~\ref{refff}.
\end{list}
\begin{subequations}\label{eq7}
\renewcommand{\theequation}{\theparentequation\alph{equation}}
\begin{equation}\label{eq7:1}
\begin{cases}
\widehat{\Delta} = \dfrac{pc-a+bN-pdM+f\widetilde{X}-pg\widetilde{Y}}{b+p^2d}\,,\\[8pt] \widehat{\Theta} =
\dfrac{hX+ \varepsilon qg\widetilde{M} - \varepsilon f\widetilde{N} - qkY}{h+q^2k}
\end{cases}
\end{equation}
\textrm{for migration from the station \fbox{I} to the station \fbox{II}\,, and for oppositely directed migration }
\begin{equation}\label{eq7:2}
\begin{cases}
\widehat{\Delta} = \dfrac{pa-c+dM-pbN+g\widetilde{Y}-pf\widetilde{X}}{d+p^2b}\,,\\[8pt] \widehat{\Theta} =
\dfrac{kY+ \varepsilon qf\widetilde{N} - \varepsilon g\widetilde{M} - qhX}{k+q^2h}\,.
\end{cases}
\end{equation}
\end{subequations}

The step~\ref{ref1111} means that the individuals of both species choose whether to emigrate, or to stay (thus determining the migration fluxes $\widehat{\Delta}$ and $\widehat{\Theta}$) taking into account the current abundances of both subpopulations and the expected abundances of the ``competing'' species evaluated due to the forthcoming migration fluxes.

\section{Results and discussion}
Reflexivity could be realized by the interacting species (within a biological community) independently. Hence, four combinations are possible, for our model:
\begin{list}{--}{\topsep=0mm \labelsep=2mm \itemsep=0mm \parsep=0mm \itemindent=3.02mm}
\item neither species realizes a reflexive strategy;
\item preys realize reflexivity in their behaviour, while predators do not;
\item predators realize reflexivity in their behaviour, but preys do not; and, finally,
\item both species realize the reflexivity in their spatial behaviour.
\end{list}
Here we consider the reflexivity in the spatial behaviour, solely. This is just a model construction, not the natural constraint. Definitely, a reflexivity can manifest through a number of issues in the life patterns and behavioural peculiarities of the individuals of some species; biologists know a lot of examples of such peculiarities~\cite{ggg,ggg1,ggg2,ggg3,ggg4,ggg5,ggg6,ggg7,ggg8}.

The impact of reflexivity on the dynamics of a community has been studied in the following manner. First of all, we define an evolution advantage through a comparison of abundances, over a sufficiently long series of generations. In other words, consider two scenarios of the community dynamics, say, the basic model (neither of the species realizes a reflexive strategy), and the scenario with preys realizing the reflexivity. Suppose, then, that all the parameters in the equations are kept the same; so that the only difference is resulted from the implementation of a reflexivity (or keeping free from that latter). Let a dynamics run in these two cases, and the abundances of each species (and moreover, in each station) are recorded. Calculate then the average figures of those abundances over a sufficiently long generation sequence (e.\,g. over~$10^4 \div 10^5$ generations). The scenario\footnote{The reflexive or reflexive-free version of a species spatial behaviour pattern is claimed to be more or less advantageous, strictly speaking.} yielding the greater average abundance figure is claimed to have an evolutionary advantage over the opposite one. To find out the situation where some specific scenario yields an advantage, we have scanned the (sub)space of parameters; see details below.

\subsection{Impact of reflexivity of preys}
\begin{figure*}[!t]
\centerline{\includegraphics[width=17cm]{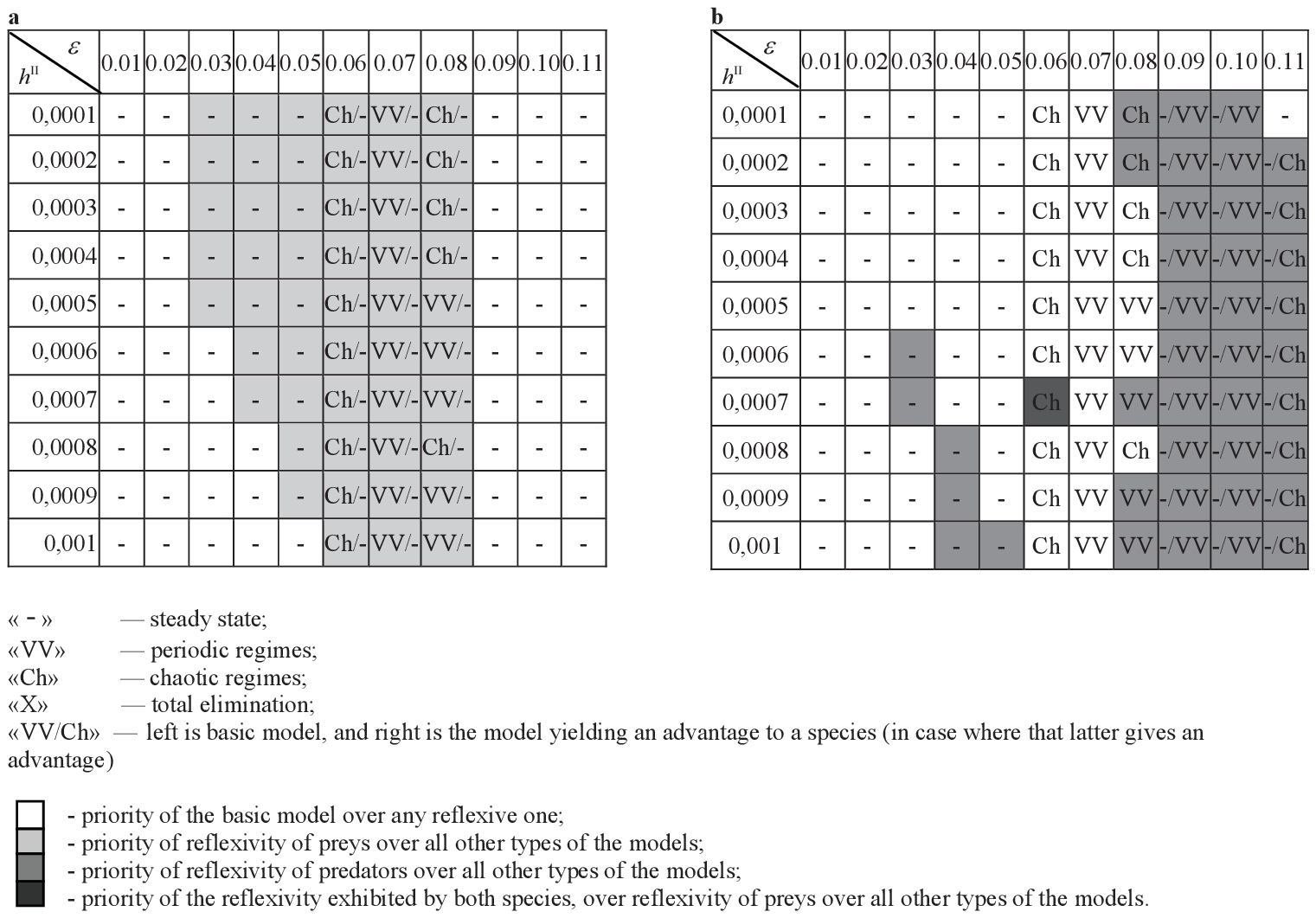}}
\caption{\label{p1} Parametric portrait of the case of comparison reflexive strategy implementation against the reflexivity-free one, for preys. The abundances of preys (a)) and predators (b)) have been measured, in dependence on $k$ and $\varepsilon$, for the fixed other parameters: $a=2$, $b=0{,}0001$, $f=0{,}0001$, $h=0{,}0002$, $c=2{,}3$, $d=0{,}0002$, $g=0{,}006$, $p=0{,}9$, $q=0{,}9$.}
\end{figure*}

Reflexivity in the preys behaviour may give them an evolutionary advantage. Fig.~\ref{p1} shows the comparison of the dynamics for two cases: reflexivity vs. non-reflexivity in the space distribution of preys. Parameters $k$ and $\varepsilon$ have been varied, while all other parameters of the systems are fixed and have the following values: $a=2$, $b=0{,}0001$, $f=0{,}0001$, $h=0{,}00002$, $c=2.3$, $d=0{,}0002$, $g=0{,}006$, $p=0{,}9$,
$q=0{,}9$. In this case, the reflexivity stabilized the dynamics of preys and resulted in the complete elimination of predators.

Evidently, the set of parameters in the figures yield an advantage to reflexive strategy bearers. The pattern seems o be rather non-random. Firstly, reflexivity gives an advantage, as a rule, to the species realizing that former. Secondly, there is a kind of ecological niches separation effect: it predators realize a reflexive behaviour, the advantage is given to those with increased assimilation coefficient. This fact may follow to a coexistence of two subspecies (both of them are predators, in our case) separated with the different strategies of spatial behaviour.

Next example illustrates the case of ``excessive'' impact of a reflexivity on the spatial behaviour. An implementation of a reflexive spatial behaviour of preys may result in a decrease of their average abundance accompanied with the growth of the average predators abundance: ``smart'' preys look more ``nutritious''. This effect may be observed with the variation of parameters $g$, $k$ and  $\varepsilon$, while the other ones are fixed at the figures $a=2{,}1$, $b=0{,}0001$, $f=0{,}002$, $h=0{,}0007$, $c=2{,}3$, $d=0{,}0002$, $p=0{,}9$, $q=0{,}9$. The remarkable fact is that for the basic model predators are eliminated, under this set of parameters. A reflexivity of preys allowed them to survive, due to a change in the dynamics (of preys) pattern. For the parameters shown above, the preys exposure a chaotic dynamics\footnote{We have no strict proof of the chaotic character of the limit dynamics of the system under consideration with the parameters set indicated above. Thus, everywhere below one should understand ``chaotic dynamics'' as that one to be complex and looking close to the chaotic one.}.

Growth of the parameters $g$, $k$ and  $\varepsilon$ causes a change of the type of a limit regimes: initially, the regime is pretty close to a periodical one, then it becomes chaotic, then again gets back to a quasi-periodic pattern. This interchange of the types of limit regimes looks quasi-periodic itself, since
\begin{figure}[!t]
\renewcommand\baselinestretch{1}
\includegraphics[width=8cm]{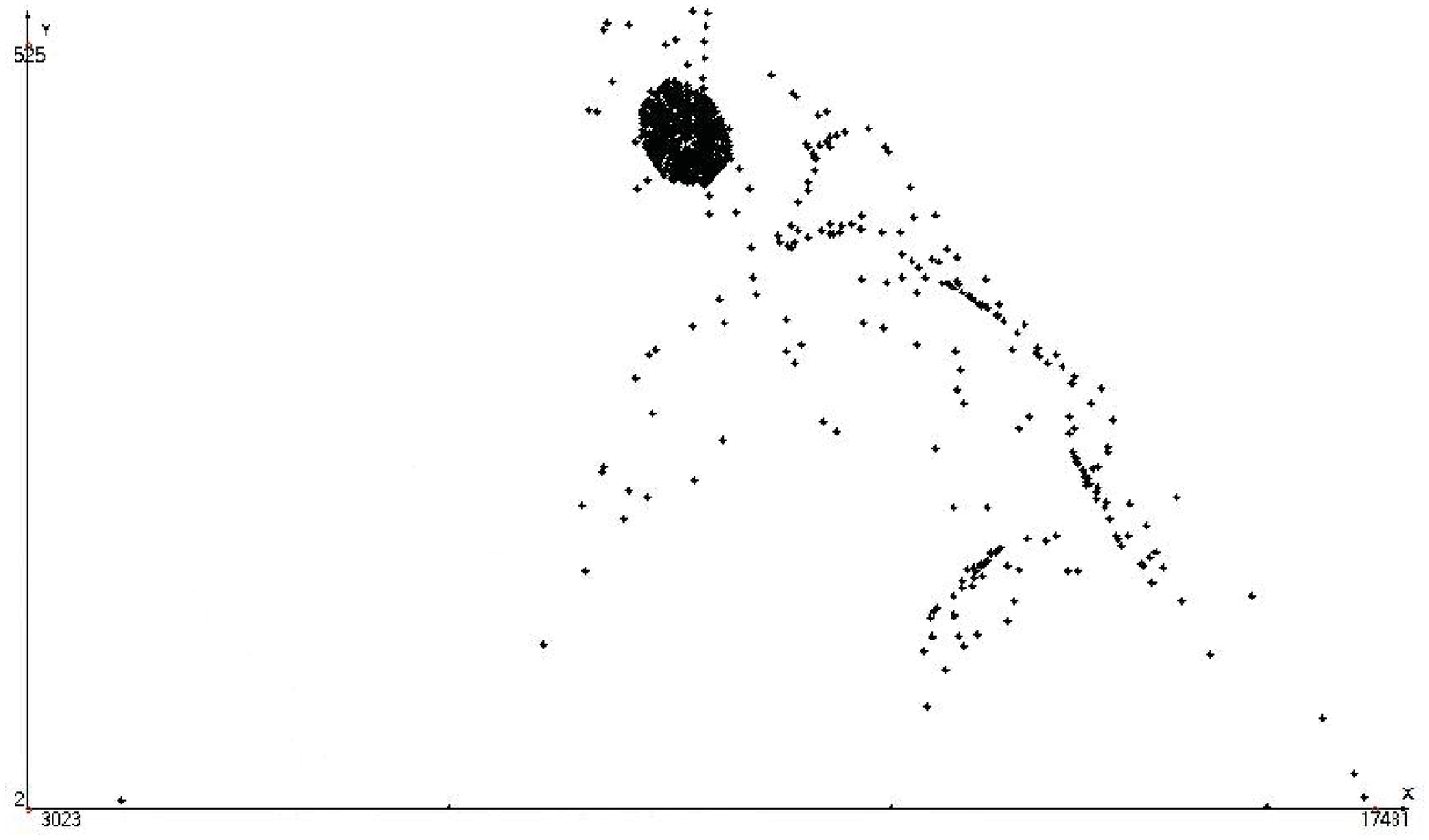}
\caption{\label{p2} Phase portrait of the dynamics of the community. Horizontal axis shows the total (over two stations) abundance of preys, and vertical axis shows that former for predators. The parameters are $a=2{,}1$, $b=0{,}0001$, $f=0{,}002$, $h=0{,}0007$, $c=2.3$, $d=0{,}0002$,
$g=0{,}002$, $k=0{,}0001$, $\varepsilon=0{,}11$, $p=0{,}9$, $q=0{,}9$.}
\end{figure}
The area in the parameters space with quasi-periodic dynamics exhibits a phase trajectory looking like \textsl{center}. Originally almost periodic dynamics changes for a chaotic one that causes a dispersion of the dots in phase plane (see Fig.~\ref{p2}). Then the dynamics comes back to a quasi-periodic pattern.

Surprisingly, reflexive behaviour of preys may bring an advantage to both species, in the model. For the fixed set of parameters $a=2$, $b=0{,}0001$, $f=0{,}002$, $h=0{,}0009$, $c=4$, $p=0{,}9$, $q=0{,}9$ with varied $d$, $g$, $k$ and $\varepsilon$, the comparison of basic model vs. that one with preys reflexivity in space distribution, one sees a growth of an average abundances of both preys and predators. Besides, an implementation of reflexivity causes a significant change in the pattern of a limit regime of dynamics.
\begin{figure}[!b]
\includegraphics[width=8cm]{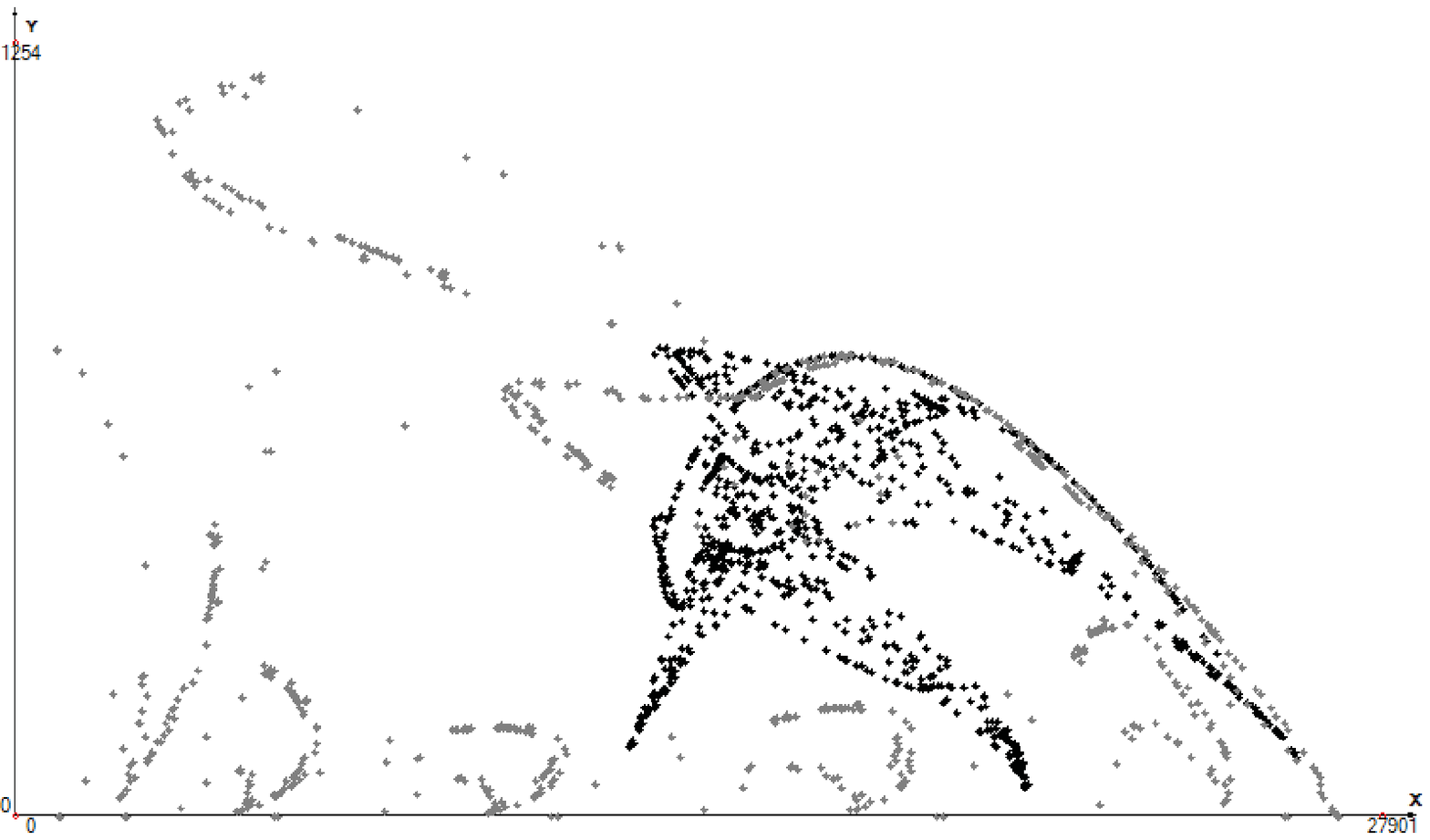}
\caption{\label{p3} Phase portrait of the dynamics of the community. Horizontal axis shows the total (over two stations) abundance of preys, and vertical axis shows that former for predators. The parameters are $a=2{,}1$, $b=0{,}0001$, $f=0{,}002$, $h=0{,}0007$, $c=2.3$, $d=0{,}0002$,
$g=0{,}002$, $k=0{,}0001$, $\varepsilon=0{,}11$, $p=0{,}9$, $q=0{,}9$.}
\end{figure}
Summarizing, one can say that for each set of the parameters shown above changes in the limit regimes have been observed, with dynamics ranged from nearly periodic to chaotic one and back. Reflexivity in the preys behaviour resulted in the decomposition of a periodic limit regime into a steady state; this decomposition followed in an increase of an average population abundance of both populations (i.\,e. preys and predators).

\subsection{Impact of reflexivity of predators}
Reflexivity implemented into the spatial behaviour of predators also may cause a growth of average abundances (interpreted as an evolutionary advantage). Let parameter $g$ and $k$ vary, while these one $a=2$, $b=0{,}0001$, $f=0{,}001$, $h=0{,}0003$, $c=1.6$, $d=0{,}0005$, $\varepsilon=0{,}11$, $p=0{,}9$, $q=0{,}9$ are fixed. Again, we scanned the plane of the parameters $g$ and $k$ with the simulation, when average abundances of both preys and predators have been calculated. The simulation has been run for basic model, and the model with reflexive behaviour of predators. For the variety of parameters $g$ and $k$, both species exhibits a steady state dynamics. Predators eliminate, for these parameters, at the basic model. Reflexivity here provided a survival of predators, thus making the abundance of that latter positive.

The set of parameters $a=2$, $b=0{,}0001$, $f=0{,}001$, $h=0{,}0003$, $c=1.6$, $d=0{,}0005$, $\varepsilon=0{,}11$, $p=0{,}9$, $q=0{,}9$ with varied figures of $d$, $g$ and $k$ provides an example where the reflexivity of predators yields the advantage to preys, simultaneously worsening the well-being of predators. This type of dynamics is observed for a significantly wide area in the space of varied parameter; Fig.~\ref{p3} shows the phase portrait of the system under the specific set of parameters. It is evident, the dynamic pattern differs significantly for two different strategies of the spatial behaviour.

The reflexivity implemented by predators solely may benefit to both species. The set of parameters $a=2$, $b=0{,}0001$, $f=0{,}003$, $h=0{,}0001$, $c=3$, $\varepsilon=0{,}08$, $p=0{,}9$, $q=0{,}9$ with varied $d$, $g$ and $k$ provide such situation.

\subsection{Combined reflexivity of both species}
\begin{figure*}
\renewcommand\baselinestretch{1}
\includegraphics[width=17cm]{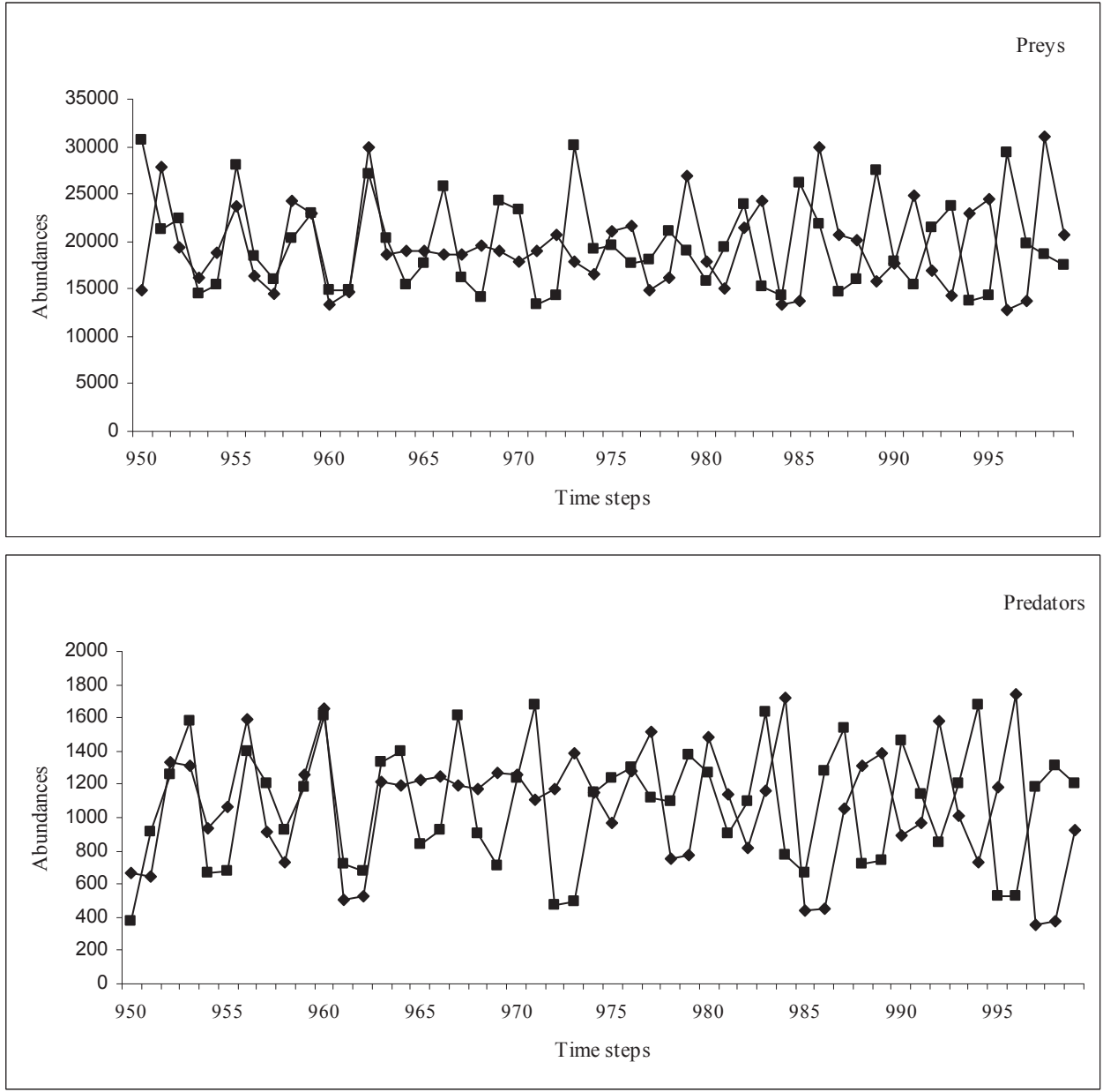}
\caption{\label{p5}  Dynamics of a community with two species reflexing each other. Horizontal axis shows the generation number, and vertical one shows the abundances. Upper chart shows the prey dynamics, and lower one shows the predator dynamics; parameters are:  $a=2$, $b=0{,}001$, $f=0{,}001$, $h=0{,}0001$, $c=4$, $d=0{,}0001$, $g=0{,}001$, $k=0{,}0005$ $\varepsilon=0{,}09$, $p=0{,}9$, $q=0{,}9$.}
\end{figure*}

A combined reflexivity in the spatial behaviour of both species may also bring the benefits to them expressed in a growth of an average abundances (of both species). Definitely, this mutual benefit is not guaranteed, and only one species may benefit from the simultaneous reflexivity. For example, simultaneous reflexivity in the behaviour of two species (within the framework of our model) brings disadvantage to preys, thus decreasing their average abundance; simultaneously, the predators increase their abundance. This regime is observed for the following parameters: $a=2$, $f=0{,}001$, $c=4$, $d=0{,}0001$, $g=0{,}001$, $k=0{,}0005$ $\varepsilon=0{,}09$, $p=0{,}9$, $q=0{,}9$. Fig.~\ref{p5} shows an example of this dynamics, for a specific set of parameters. Remarkable fact is that the type of the dynamics (chaotic-like pattern) is observed in both models (i.\,e. in basic one, and the reflexive model).

\section{Conclusion}
An issue of the targeted (non-diffuse, etc.) migration is heavily dependent on very important idea that is information access and knowledge available to make a decision. Both these ideas are hard to define and hard to study. Definitely, very few animals may exhibit a behaviour based on ``knowledge'' that is similar, to some extent, to that one peculiar for human beings. Nonetheless, a tremendous number of examples make an evidence of a non-random and reasonable behaviour of the beings of extremely different taxa. Thus, observations over animals contribute a lot the methodology of finite automata \cite{ggg12,ggg15,ggg16}. This approach forces to figure out what is the information necessary to act (in some proper way), exactly.

In general, such information could be separated into two (generalized) issues: the former is related to an outer, external situation, and the latter describes the status of a being (say, hunger, experience, pain, etc.; see, e.\,g., \cite{ggg7,ggg8,ggg10}). An external information concerns, first of all, the environmental conditions of a site: local population density is number one among them \cite{ggg2,ggg3,ggg4,ggg7,ggg8,ggg13}. Unlike a general approach in modelling, where an average population density is used to model (or simulate) the dynamics of a community, the reality is based on the local density figures.

Yet, it is not the end of a problem. The information used by an individual to evaluate a situation and make a decision to realize some specific behavioural act significantly differs in the value of a specific radius of access of that former. Here the problem of a scale arises; there are two natural ways to determine that latter:
\begin{list}{--}{\topsep=0mm \labelsep=2mm \itemsep=0mm \parsep=0mm \itemindent=.02mm}
\item a body scale distance; and
\item an individual specific transfer distance.
\end{list}
So, an effective local population density should be determined through a comparison of the traces of a number of individuals. The ways to identify those traces are numerous and extremely diverse. Basically, they could be arranged into three groups. The most widespread and effective way is a chemical communication (smell); some media do not allow such way of communication, and rather advanced techniques have been elabourated \cite{ggg13}. The species occupying extended territories use visual labelling.

In other words, some information may come in diverse ways, and from different distances. Probably, different kinds of the information have different scales. Apparently, a communication related to reproduction behaviour has the largest scale. Reciprocally, a trophic behaviour patterns are at the opposite pole of the scale. Finally, there are socially determined effects in the problem of information accessibility and scaling: reflexivity caused by a competitive (with neither respect to a nature of this competition) origin of the behavioural patterns (see, e.\,g., \cite{ggg14}

Simulation shows that reflexivity of either species impacts both the average abundances, and the dynamics type of populations in the community. Chaotic dynamics may exhibit a variation in the average abundances of both species, with no change in the type of the dynamics. For periodic limit regimes, an implementation of reflexivity may cause a phase shift, in the dynamics: the oscillations of an abundance of preys realizing reflexive strategy run ``ahead'' of those observed in the basic model. A change of a limit regime type in case of reflexivity implementation are also related to the dependence of a limit regime type on the initial data. Of course, this ``dependence'' means the significant change of an attractive set specific for the given limit regime type, in the space of initial conditions. Such effect has been observed for chaotic dynamics, only.

An advantage provided by the implementation of a reflexivity (either for preys, or for predators) makes no surprise. The cases where realization of reflexivity by one species provided an advantage for another one simultaneously deteriorating the survival of the given one seem to be more intriguing. Meanwhile, such effect is quite rare: an improvement of the living conditions for another species usually does not deteriorate the living conditions of the given one. Migration is the necessary, while not sufficient condition of the improvement resulted from the reflexivity: that former is the only way to realize a reflexive strategy.

Reflexivity may result in the migration occurrence, when that latter was absent in the basic (reflection-free) model. Nonetheless, reflexivity does not increase adaptivity of a system always. As we saw, reflexivity might have no effect on the dynamics of a system, at all, or even deteriorate the living conditions of the species realizing the reflexive behaviour. Whether reflexivity gives an advantage, or not, depends strongly on the specific figures of the parameters of the system under consideration. A detailed structure of a parametric portrait must be studied specially. In case, when reflexivity does not bring any advantage to the species realizing that, one may see the following reasons for that:
\begin{list} {--} {\topsep=0mm \labelsep=2mm \itemsep=0mm \parsep=0mm \itemindent=3.02mm}
\item there is no way to increase an average abundance, either with reflexivity, or without it, at the given conditions (``super-optimal'' environmental conditions);
\item global information access provides too rigid rule to determine a migration flux: ``excessive knowledge'' kills adaptivity.
\end{list}

Approach developed in this paper could be extended for the study of the biological communities incorporating three and more species related with a variety of interactions, including \textsl{prey~-- predator} relation. Another types of interacting agents could be easily implemented into a model of this kind. Besides, basically new type of information access pattern could be implemented within the framework of such models; that is a local information access.
\end{document}